\documentclass[%
 reprint,
superscriptaddress,
 amsmath,amssymb,
 aps,
]{revtex4-2}
\usepackage{lipsum}
\usepackage{graphicx}
\usepackage{dcolumn}
\usepackage{bm}
\usepackage{hyperref}
\usepackage{braket}
\usepackage{xcolor}
\graphicspath{{figures/}}

\begin{document}


\title{Simulating rare kaon decays \texorpdfstring{$K^{+}\to\pi^{+}\ell^{+}\ell^{-}$}{KtoPiEllEll}\\using domain wall lattice QCD with physical light quark masses}

\author{P. A. Boyle}
\affiliation{School of Physics and Astronomy, University of Edinburgh, Edinburgh EH9 3JZ, UK}
\affiliation{Physics Department, Brookhaven National Laboratory, Upton NY 11973, USA}

\author{F. Erben}
\affiliation{School of Physics and Astronomy, University of Edinburgh, Edinburgh EH9 3JZ, UK}

\author{J. M. Flynn}
\affiliation{School of Physics and Astronomy, University of Southampton, Southampton, SO17 1BJ, UK}
\affiliation{STAG Research Centre, University of Southampton, Southampton, SO17 1BJ, UK}

\author{V. G\"ulpers}
\affiliation{School of Physics and Astronomy, University of Edinburgh, Edinburgh EH9 3JZ, UK}

\author{R. C. Hill}
\affiliation{School of Physics and Astronomy, University of Edinburgh, Edinburgh EH9 3JZ, UK}
\affiliation{School of Physics and Astronomy, University of Southampton, Southampton, SO17 1BJ, UK}

\author{R.~Hodgson}
\affiliation{School of Physics and Astronomy, University of Edinburgh, Edinburgh EH9 3JZ, UK}

\author{A. J\"uttner}
\affiliation{School of Physics and Astronomy, University of Southampton, Southampton, SO17 1BJ, UK}
\affiliation{STAG Research Centre, University of Southampton, Southampton, SO17 1BJ, UK}
\affiliation{CERN, Theoretical Physics Department, Geneva, Switzerland}

\author{F. \'O h\'Og\'ain}
\affiliation{School of Physics and Astronomy, University of Edinburgh, Edinburgh EH9 3JZ, UK}

\author{A. Portelli}
\email{antonin.portelli@ed.ac.uk}
\affiliation{School of Physics and Astronomy, University of Edinburgh, Edinburgh EH9 3JZ, UK}

\author{C. T. Sachrajda}
\affiliation{School of Physics and Astronomy, University of Southampton, Southampton, SO17 1BJ, UK}

\collaboration{RBC and UKQCD Collaborations}

\date{\today}

\begin{abstract}
We report the first calculation using physical light-quark masses of
the electromagnetic form factor $V(z)$ describing the long-distance
contributions to the $K^+\to\pi^+\ell^+\ell^-$ decay amplitude. The
calculation is performed on a 2+1 flavor domain wall fermion ensemble
with inverse lattice spacing $a^{-1}=1.730(4)$GeV. We implement a
Glashow-Iliopoulos-Maiani cancellation by extrapolating to the
physical charm-quark mass from three below-charm masses. We obtain
$V(z=0.013(2))=-0.87(4.44)$, achieving a bound for the value. The
large statistical error arises from stochastically estimated quark loops.
\end{abstract}
\maketitle

\section{Introduction}
\label{sec:introduction}
The $K^{+}\to\pi^{+}\ell^{+}\ell^{-}$ ($\ell=e,\mu$) decays are flavor-changing neutral current processes that are heavily suppressed in the standard model (SM), and thus expected to be sensitive to new physics. Their branching ratios, taken from the latest PDG average~\cite{Zyla:2020zbs}, are $\text{Br}\left[K^{+}\to\pi^{+} e^{+} e^{-}\right] = 3.00(9)\times10^{-7}$ and  $\text{Br}\left[K^{+}\to\pi^{+}\mu^{+}\mu^{-}\right] = 9.4(6)\times10^{-8}$. This process is dominated by a single virtual-photon exchange $\left(K\to\pi\gamma^{*}\right)$, whose amplitude is predominantly described by long-distance, nonperturbative physics~\cite{D_Ambrosio_1998}.
With tensions between the LHCb measurement~\cite{LHCb:2021trn} of and SM predictions for the ratio $R_K$ contributing to increased interest in lepton-flavor universality (LFU) violation, important tests of LFU in the kaon sector could also be provided by $K^+\to\pi^+\ell^+\ell^-$ decays~\cite{Crivellin_2016}.
\indent The amplitude for the $K\to\pi\gamma^{*}$ decay can be expressed in terms of a single electromagnetic form factor $V(z)$ defined \textit{via} \cite{D_Ambrosio_1998,RevModPhys.84.399}
\begin{equation}
    \label{eq:amplitude_defn}
    \mathcal{A}_{\mu} = -i\frac{G_{F}}{(4\pi)^{2}}V(z)[q^{2}\left(k+p\right)_{\mu} - (M_{K}^{2} - M_{\pi}^{2})q_{\mu}],
\end{equation}
where $\mu$ is the photon polarisation index, $z=q^{2}/M_{K}^{2}$, $q=k-p$, and $k$ and $p$ indicate the momenta of the $K$ and $\pi$ respectively. From analyticity, a prediction of $V(z)$ is given by \cite{D_Ambrosio_1998}
\begin{equation}
    \label{eq:form_factor_defn}
    V(z) = a_{+} + b_{+}z + V^{\pi\pi}(z),
\end{equation}
where $a_{+}$ and $b_{+}$ are free real parameters and $V^{\pi\pi}(z)$ describes the contribution from a $\pi\pi$ intermediate state (detailed  in~\cite{D_Ambrosio_1998}) with a $\pi^{+}\pi^{-}\to\gamma^{*}$ transition. The free parameters have, until recently, only been obtained by fitting experimental data. Having previously measured the $K^{+}$ decay channel for electrons and muons at the NA48 experiment at the CERN SPS \cite{BATLEY2009246}, the follow-up NA62 experiment measured the $K^{+}\to\pi^{+}\mu^{+}\mu^{-}$ decay during the 2016-2018 Run 1 \cite{Bician:2021WP}, with prospects for further measurements during the 2021-2024 Run 2 \cite{2021NA62report}. From the NA48 electron data, values of $a_{+}=-0.578(16)$ and $b_{+}=-0.779(66)$ have been found \cite{BATLEY2009246}, and the available NA62 muon data resulted in $a_{+}=-0.592(15)$ and $b_{+}=-0.699(58)$ \cite{Bician:2021WP}. \\
\indent In parallel, the theoretical understanding of these processes is being improved. The authors of \cite{D_Ambrosio_2019,D_Ambrosio_2019b} construct a theoretical prediction of $a_{+}$ and $b_{+}$ by considering a two-loop low-energy expansion of $V(z)$ in three-flavor QCD, with a phenomenological determination of quantities unknown at vanishing momentum transfer. From the electron and muon they find $a_{+}=-1.59(8)$ and $b_{+}=-0.82(6)$, in significant tension with the experimental data fit. The authors acknowledge that more work is being done to estimate more accurately the $\pi\pi$ and $KK$ contributions. \\
\indent The nonperturbative \textit{ab-initio} approach of lattice QCD is well suited to study the dominant long-distance contribution to the matrix element of the $K^{+}\to\pi^{+}\gamma^{*}$ decay. Methods with which such a lattice calculation could be performed were first proposed in \cite{ISIDORI200675}, and additional details on full control of ultraviolet divergences were introduced in \cite{Christ_2015}. An exploratory lattice calculation  \cite{PhysRevD.94.114516}, using unphysical meson masses, demonstrated a practical application of these methods.\\
\indent This letter describes a lattice calculation following the same approach as \cite{PhysRevD.94.114516}, but using physical light-quark masses, thereby allowing for the first time a direct comparison to experiment.
\section{Extraction of the Decay Amplitude}
\label{sec:determination_of_amplitude}
The procedure and expressions in this section are largely a summary of the approach described in~\cite{PhysRevD.94.114516}.
We wish to compute the long-distance amplitude defined as
\begin{equation}
    \label{eq:Minkowski_amplitude}
    \mathcal{A}_{\mu}\left(q^{2}\right)=\int d^{4}x\bra{ \pi(\mathbf{p})}T\left[J_{\mu}(0)H_{W}(x)\right]\ket{K(\mathbf{k})}
\end{equation}
in Minkowski space, where $q$, $k$, and $p$ are defined as above, $J_{\mu}$ is the quark electromagnetic current and $H_{W}$ is a $\Delta S=1$ effective Hamiltonian density, given by \cite{RevModPhys.68.1125}
\begin{equation}
    \label{eq:weak_hamiltonian}
    H_{W}=\dfrac{G_{F}}{\sqrt{2}}V_{us}^{*}V_{ud}\,\sum_{j=1}^{2}C_{j}\left(Q_{j}^{u}-Q_{j}^{c}\right),
\end{equation}
where the $C_{j}$ are Wilson coefficients, and $Q_{1}^{q}$ and $Q_{2}^{q}$ are the current-current operators defined (up to a Fierz transformation) by \cite{ISIDORI200675}
\begin{align}
    \label{eq:Q1_Q2_definition}
    Q_{1}^{q}&=[\bar{s}\gamma_{\mu}\left(1-\gamma_{5}\right)d][\bar{q}\gamma^{\mu}\left(1-\gamma_{5}\right)q]\,, \\%
    Q_{2}^{q}&=[\bar{s}\gamma_{\mu}\left(1-\gamma_{5}\right)q][\bar{q}\gamma^{\mu}\left(1-\gamma_{5}\right)d]\,.
\label{eq:one}
\end{align}
We renormalize the operators $Q^{q}_{i}$ nonperturbatively within the RI-SMOM scheme~\cite{Sturm:2009kb} and then follow \cite{PhysRevD.84.014001} to match to the $\overline{\mathrm{MS}}$ scheme, in which the Wilson coefficients have also been computed.
\subsection{Correlators and Contractions}
\label{sub:Contractions}

The corresponding Euclidean amplitude---which is accessible to lattice QCD calculations---can be computed with the ``unintegrated'' 4pt correlator~\cite{Christ_2015}
\begin{multline}
    \label{eq:4pt_corr}
    \Gamma_{\mu}^{\left(4\right)}\left(t_{H},t_{J},\mathbf{k},\mathbf{p}\right)=\int d^3\mathbf{x}\int d^3\mathbf{y}\,e^{-i\mathbf{q}\cdotp\mathbf{x}}\\
    \langle \phi_{\pi}\left(t_{\pi},\mathbf{p}\right)T\left[J_{\mu}\left(t_{J},\mathbf{x}\right)O_{W}\left(t_{H},\mathbf{y}\right)\right]\phi_{K}^{\dagger}\left(t_K,\mathbf{k}\right)\rangle,
\end{multline}
where $\phi_{P}^{\dagger}\left(t,\mathbf{k}\right)$ is the creation operator for a pseudoscalar meson $P$ at time $t$ with momentum $\mathbf{k}$. To obtain the decay amplitude we take the integrated 4pt correlator~\cite{Christ_2015}
\begin{multline}
    \label{eq:integrated_4pt_corr}
    I_{\mu}\left(T_{a},T_{b},\mathbf{k},\mathbf{p}\right)=e^{-\left(E_{\pi}\left(\mathbf{p}\right)-E_{K}\left(\mathbf{k}\right)\right)t_{J}} \\
    \times \int_{t_{J}-T_{a}}^{t_{J}+T_{b}}dt_H\,\tilde{\Gamma}_{\mu}^{\left(4\right)}\left(t_{H},t_{J},\mathbf{k},\mathbf{p}\right),
\end{multline}
in the limit $T_{a},T_{b}\to\infty$. The exponential factor translates the decay to $t_{J}=0$, allowing us to omit any $t_{J}$ dependence in further expressions. Here $\tilde{\Gamma}_{\mu}^{\left(4\right)}$ is the ``reduced'' correlator, where we have divided out factors that are not included in the final amplitude, \textit{i.e}.
\begin{equation}
    \label{eq:reduced_4pt_corr}
    \tilde{\Gamma}_{\mu}^{\left(4\right)}=\dfrac{\Gamma_{\mu}^{\left(4\right)}}{Z_{\pi K}},\quad Z_{\pi K}=\dfrac{Z_{\pi}Z_{K}^{\dagger}L^3}{4E_{\pi}\left(\mathbf{p}\right)E_{K}\left(\mathbf{k}\right)}e^{-t_{\pi}E_{\pi}\left(\mathbf{p}\right)+t_{K}E_{K}\left(\mathbf{k}\right)},
\end{equation}
where $L^{3}$ is the spatial volume, $Z_{\pi}=\bra{ 0 }\phi_{\pi}(\mathbf{p})\ket{\pi(\mathbf{p})}$, $Z^{\dagger}_{K}=\bra{K(\mathbf{k})}\phi_{K}^{\dagger}(\mathbf{k})\ket{0}$, and $E_K(\mathbf{k})$ and $E_{\pi}(\mathbf{p})$ are the initial-state kaon and final-state pion energies, respectively.\\ 
\indent The spectral decomposition of Eq.~(\ref{eq:4pt_corr}) has been discussed in detail in \cite{PhysRevD.94.114516}, in particular describing the presence of intermediate one-, two-, and three-pion states between the $J_{\mu}$ and $O_{W}$ operators. As these states can have energies $E<E_{K}(\mathbf{k})$ they introduce exponentially growing contributions that cause the integral to diverge with increasing $T_{a}$. These contributions do not contribute to the Minkowski decay width~\cite{Christ_2015} and must be removed in order to extract the amplitude
\begin{equation}
    \label{eq:Minkowski_amp_from_4pt_int}
    A_{\mu}(q^2) = \lim_{T_{a},T_{b}\rightarrow\infty}\tilde{I}_{\mu}(T_{a},T_{b},\mathbf{k},\mathbf{p}),
\end{equation}
where $\tilde{I}_{\mu}$ is the integrated 4pt correlator with intermediate-state contributions subtracted. 
The methods used to remove the intermediate states follow the same steps as in \cite{PhysRevD.94.114516}, and are outlined in Section~\ref{sub:intermediate_states}.\\
\indent The four classes of diagrams---Connected ($C$), Wing ($W$), Saucer ($S$), and Eye ($E$)---that contribute to the integrated correlator are represented schematically in the supplementary material. The current can be inserted on all four quark propagators in each class of diagram, in addition to a quark-disconnected self-contraction. Diagrams of these five current insertions for the $C$ class are also shown in the supplementary material. The 20 resulting diagrams need to be computed in order to evaluate Eq.~(\ref{eq:4pt_corr}).\\
\indent When working on the lattice there are potentially quadratically divergent contributions that come about as the operators $J_{\mu}$ and $H_{W}$ approach each other when the current is inserted on the loop of the $S$ and $E$ diagrams \cite{Christ_2015, ISIDORI200675}. Since we perform our calculation with conserved electromagnetic currents the degree of divergence is reduced to, at most, a logarithmic divergence \cite{ISIDORI200675} as a consequence of $\mathrm{U}(1)$ gauge invariance and the resulting Ward-Takahashi identity. We emphasise that, due to exact gauge symmetry in lattice QCD there is a vector current, which is exactly conserved on each configuration, independent of any residual chiral symmetry breaking. The remaining logarithmic divergence is removed through the Glashow-Iliopoulos-Maiani (GIM) mechanism \cite{PhysRevD.2.1285}, implemented here through the inclusion of a valence charm quark in the lattice calculation.

\subsection{Intermediate states}
\label{sub:intermediate_states}
The contribution of the single-pion intermediate state can be removed by either of the two methods discussed in \cite{PhysRevD.94.114516}. The first of these (method 1) reconstructs the single-pion state using 2pt and 3pt correlators to subtract its contribution explicitly. The relevant amplitude can be extracted with this method in several ways, including a direct fit of $A_{\mu}$ and the intermediate state, the reconstruction of the intermediate states using fits to 2pt and 3pt correlators, a zero-momentum-transfer approximation and an $\mathrm{SU}(3)$-symmetric-limit approximation, all of which are discussed in detail in \cite{PhysRevD.94.114516}.\\
\indent The second method proposed in \cite{PhysRevD.94.114516} (method 2) involves an additive shift to the weak Hamiltonian by the scalar density $\bar{s}d$ \cite{PhysRevLett.113.112003}
\begin{equation}
    \label{eq:sd_shift}
    O_{W}^{\prime} = O_{W}-c_{s}\bar{s}d,
\end{equation}
where the constant parameter $c_{s}$ is  chosen such that
\begin{equation}
    \label{eq:cs_defn}
    \bra{\pi(\mathbf{k})}O_{W}^{\prime}\ket{K(\mathbf{k})} = 0.
\end{equation}
Replacing $O_{W}$ with $O_{W}^{\prime}$ in Eq.~(\ref{eq:4pt_corr}) removes the contribution of the single-pion intermediate state. As the scalar density can be written in terms of the divergence of a current, the physical amplitude is invariant under such translation~\cite{Christ_2015}.
The two-pion contributions are expected to be insignificant until calculations reach percent-level precision and the three-pion states are even more suppressed \cite{Christ_2015}. As we do not compute the rare kaon decay amplitude to such a precision, the two- and three-pion states are not accounted for in our studies.

\section{Details of Calculation}
This calculation is performed on a lattice ensemble generated with the Iwasaki gauge action and 2+1 flavors of M\"obius domain wall fermions (DWF) \cite{PhysRevD.93.074505}. The spacetime volume is $(L/a)^3 \times (T/a) = 48^3\times96$ and the inverse lattice spacing $a^{-1}=1.730\left(4\right)$GeV. The fifth-dimensional extent is $L_{s}=24$ and the residual mass is $am_{\text{res}}=6.102\left(40\right)\times10^{-4}$. The light and strange sea quark masses are $am_{l} = 0.00078$ and $am_{s}=0.0362$ respectively, corresponding to  pion and kaon masses of $M_{\pi}=139.2(4)$MeV and $M_{K}=499(1)$MeV. We use 87 gauge configurations, each separated by 20 Monte 
Carlo time steps.

The M\"obius DWF action \cite{BROWER20171} was used to simulate the sea quarks, with a rational approximation used for the strange quark. In this calculation the light valence quarks make use of the zM\"obius action \cite{McGlynn:2016OM}, an approximation of the M\"obius action where the sign function has had its $L_{s}$ dimension reduced by using complex parameters matched to the original real parameters using the Remez algorithm. This gives a reduced fifth-dimensional extent $L_{s}=10$, reducing the computational cost of light-quark inversions. The lowest $2000$ eigenvectors of the Dirac operator were also calculated (``deflation''), allowing us to accelerate the light-quark zM\"obius inversions further. We correct for the bias introduced by the zM\"obius action with a technique similar to all-mode-averaging (AMA) \cite{PhysRevD.88.094503} by computing light and charm propagators also using the M\"obius action on lower statistics, using the M\"obius accelerated DWF (MADWF) algorithm \cite{Yin:2011np} with deflated zM\"obius guesses in the inner loop of the algorithm for the light and a mixed-precision solver for the charm quarks. Further details are in the supplementary materials.

The GIM subtraction relies on a precise cancellation, in particular in the low modes of the light and charm actions, and it is paramount to use the same actions for those quarks. With the choice of zM\"obius parameters for the light quark, the DWF theory breaks down for the physical charm-quark mass~\cite{Boyle_2016}. We instead perform the GIM subtractions using three unphysical charm-quark masses, chosen to be $am_{c_{1}}=0.25$, $am_{c_{2}}=0.30$, $am_{c_{3}}=0.35$,  and extrapolate the results to the physical point. The physical charm-quark mass was found to be $am_{c}=0.510(1)$ by computing the three unphysical $\eta_{c}$-meson masses and extrapolating to the physical $\eta_{c}$ mass. Previous work has demonstrated that, for the lattice parameters in use for this calculation, such an extrapolation is well-controlled~\cite{DDsDecayConstantsRBCUKQCD}.

We use Coulomb-gauge fixed wall sources for the kaon and pion. The pion and kaon sources are separated by 32 lattice units in time, with the kaon at rest at $t_{K}=0$ and the pion with momentum $\mathbf{p}=\frac{2\pi}{L}\left(1,0,0\right)$ at $t_{\pi}=32$. The electromagnetic current is inserted midway between the kaon and the pion at $t_{J}=16$, so that the effects of the excited states from the interpolating operators are suppressed.
We omit the disconnected diagram, since it is suppressed by $\mathrm{SU}(3)$ flavour symmetry and $1/N_c$ to an expected $\sim 10\%$ of the connected-diagram contribution~\cite{PhysRevD.94.114516}. Given the error on our final result, the disconnected contribution is negligible. Control of the error is being explored in an ongoing project.

We use the M\"obius conserved lattice vector current \cite{PhysRevD.93.074505} with only the time component $\mu=0$, which is sufficient to extract the single form factor from Eq.~(\ref{eq:amplitude_defn}).

\begin{figure*}
    \begin{centering}
        \includegraphics[scale=1]{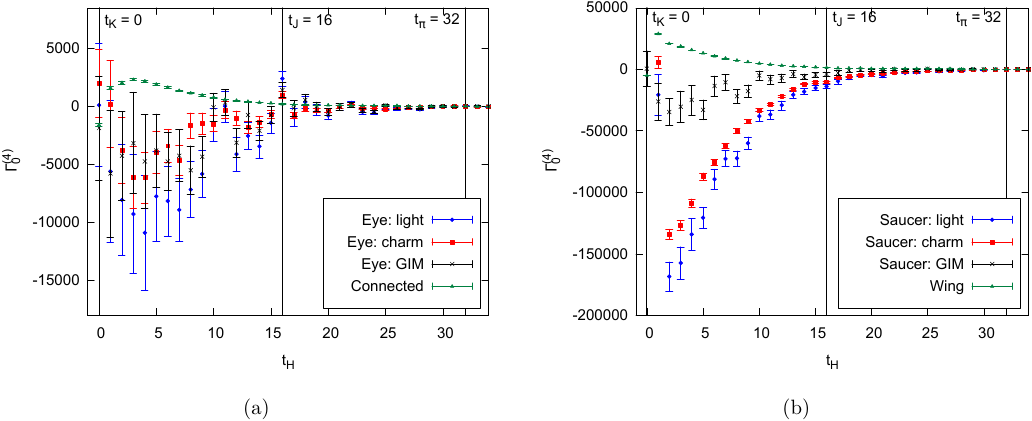}
    \par\end{centering}
\protect\caption{\label{fig:Q1Q2_c1}The (a) $Q_{1}$ and (b) $Q_{2}$ operator contributions to the $am_{c_{1}}=0.25$ integrated 4pt rare kaon correlator, separated into $C$, $W$, $S$, and $E$ diagrams. The light- and charm-quark contributions to the $S$ and $E$ diagrams are shown individually, as well as their difference, the ``GIM'' contribution.}
\end{figure*}
To compute the loops in the $S$ and $E$ diagrams we use spin-color diluted sparse sources, similar to those used in \cite{PhysRevLett.116.232002}, the structure of which is described in the supplemental material. We use the AMA technique \cite{PhysRevD.88.094503} for our calculation of these diagrams, computing one hit of sparse noise with ``exact'' solver precision ($10^{-8}$, $10^{-10}$, $10^{-12}$, and $10^{-14}$ for the light, $c_{1}$, $c_{2}$, and $c_{3}$ quarks, respectively) and the same hit of sparse noise with ``inexact'' solver precision ($10^{-4}$ for all quarks). We then compute an additional 9 hits of sparse noise with inexact solver precision and apply a correction computed from the difference of the reciprocal noises.

We performed all correlation function calculations using dedicated software~\cite{fionn_o_hogain_2022_6369186} based on the Grid~\cite{Boyle:2016lbp, Boyle:2022nef} and Hadrons~\cite{antonin_portelli_2022_6382460} libraries. All three are free software under GPLv2. The raw lattice correlators used in this work are publicly available online~\cite{peter_a_boyle_2022_6369178}.

\section{Numerical Results}
\begin{figure}[t!]
	\begin{centering}
		\includegraphics[scale=0.28]{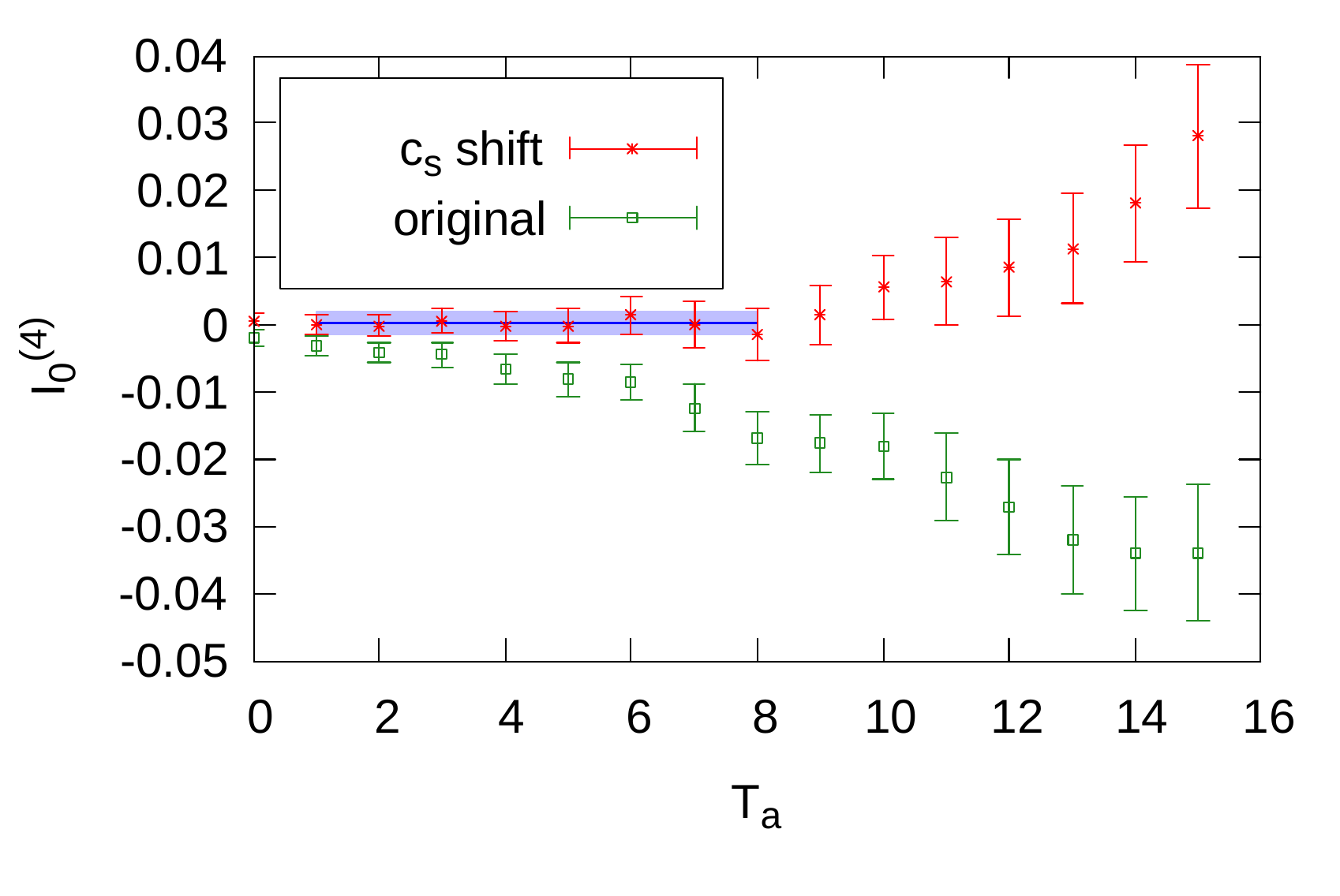}
	\end{centering}
	\caption{\label{fig:4pt_int_corr_result} The $am_{c_{1}}=0.25$ integrated 4pt rare kaon correlator shown for $I_{0}\left(T_{a}, T_{b} = 8, \mathbf{k}, \mathbf{p}\right)$ (cf. Eq.~(\ref{eq:integrated_4pt_corr})) to demonstrate the $T_{a}$ dependence. The green data shows the raw 4pt function, and in red we show the same data after removing the single-pion exponential growth \textit{via} method 2. The fit to the plateau, shown in blue, gives $A_{0} = 0.00022(172)$.}
\end{figure}

The 4pt functions for the lightest charm-quark mass are shown in Fig.~\ref{fig:Q1Q2_c1}, and Fig.~\ref{fig:4pt_int_corr_result} shows the $T_{a}$ dependence of the integrated correlator for fixed $T_{b}$ both before and after removing the exponentially growing contributions using method 2. We perform a simultaneous fit to the 2pt, 3pt and integrated 4pt functions, extracting matrix elements, energies, form factors and $A_{0}$, using a covariance matrix with fully correlated 2pt and 3pt sectors and uncorrelated 4pt sector. From this fit, we obtain $A_{0} = 0.00022(172)$ with a $\chi^2/\text{dof}=0.996$. Further details on the fitting procedure, including a discussion of the fit ranges which were used, are presented in the supplemental material. The error on $A_{0}$ is entirely statistical.

\begin{figure}[t!]
	\begin{centering}
		\includegraphics[scale=0.33]{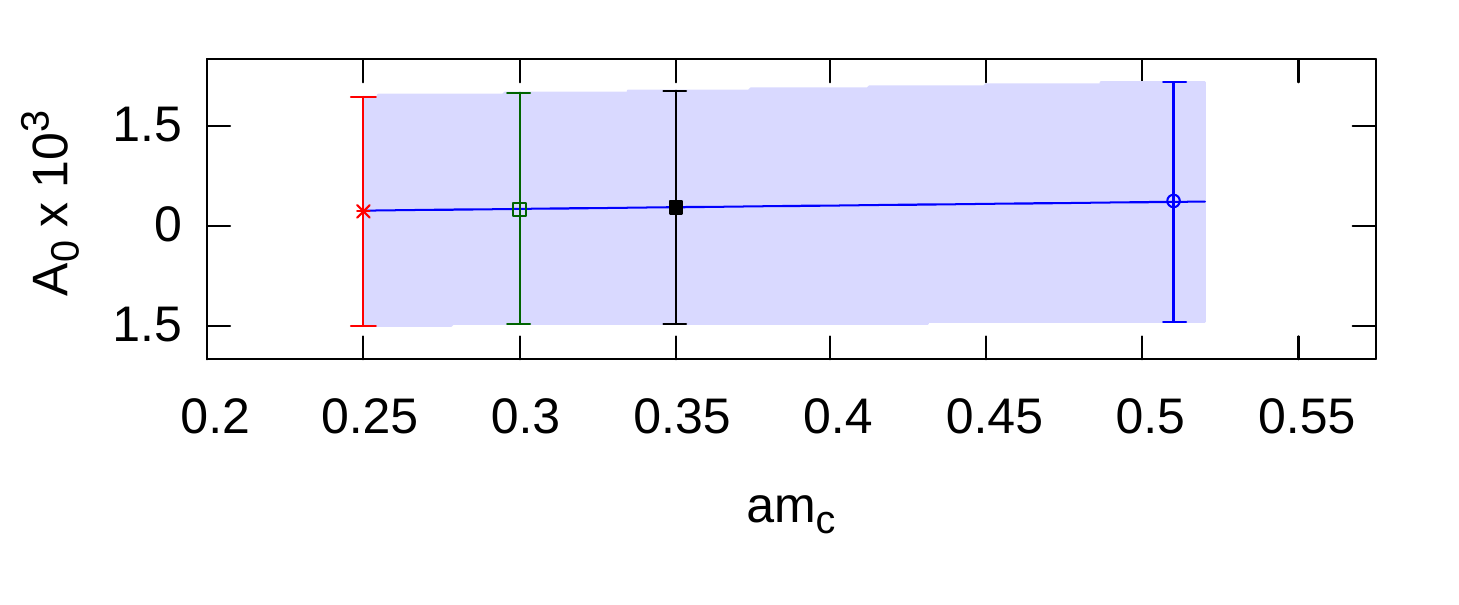}
	\end{centering}
	\caption{\label{fig:Mc_extrapolated_A0} The extrapolation of the $A_{0}$ results found using method 2 to the physical charm-quark mass. The linear fit and extrapolated result are shown in blue, giving a result of $A_{0}=0.00035(180)$. Red, green, and black show the results at the $c$ masses we simulate at, and we extrapolate those to the blue data point at physical charm mass.}
\end{figure}

Table~\ref{tab:A0_results} shows the results for $A_{0}$ using the three charm-quark masses, extracted using the different methods detailed above. The results from method 2 have statistical errors compatible with method 1 results. As method 2 has the simplest fit structure, we use it to extrapolate to the physical charm-quark mass and to compute the form factor as our final result. We stress that method 1 remains an important cross-check on the analysis.
\begin{table}[b]
\caption{\label{tab:A0_results}%
Fit results for $A_{0}$ for the three unphysical charm-quark masses and value found from extrapolating these to the physical point. The first four results are obtained using the various approaches to method 1, as described in Section~\ref{sub:intermediate_states}, and the final result is obtained using method 2.
}
\begin{ruledtabular}
\begin{tabular}{lccc}
\textrm{Analysis}&
\textrm{$m_{c_{1}}$}&
\textrm{$m_{c_{2}}$}&
\textrm{$m_{c_{3}}$}\\
\colrule
\hline
\textit{Method 1} & & & \\
\hline
 Direct fit & -0.00052(208)  & -0.00046(210)  & -0.00040(211) \\
\hline
 2pt/3pt recon & -0.00036(162) & -0.00024(164) & -0.00017(165) \\
\hline
 0 mom transfer & -0.00087(165) & -0.00086(166) & -0.00086(167) \\
\hline
 $SU(3)$ symm lim & 0.00055(165)  & 0.00085(166)  & 0.00112(167) \\
\hline
\textit{Method 2} & & & \\
\hline
 $c_s$ shift & 0.00022(172) & 0.00024(173) & 0.00027(174) \\
\end{tabular}
\end{ruledtabular}
\end{table}
Fig.~\ref{fig:Mc_extrapolated_A0} shows the extrapolation of the method-2 results to the physical charm-quark mass, giving a value of $A_{0}=0.00035(180)$. From Eq.~(\ref{eq:amplitude_defn}) we can relate our result to the form factor to achieve $V\left(z\right)=-0.87(4.44)$. For our choice of kinematics we have $z=0.013(2)$; we expect the $b_+z$ contribution to be $\sim 10^{-2}$ assuming $b_+$ is $\mathcal{O}(1)$, and we estimate $V^{\pi\pi}(z) = -0.00076(73)$ following \cite{D_Ambrosio_1998}. We may therefore take our result for $a_{+}$ as an approximation for the intercept of the form factor.
\begin{figure*}[t!]
	\begin{centering}
		\includegraphics[scale=1]{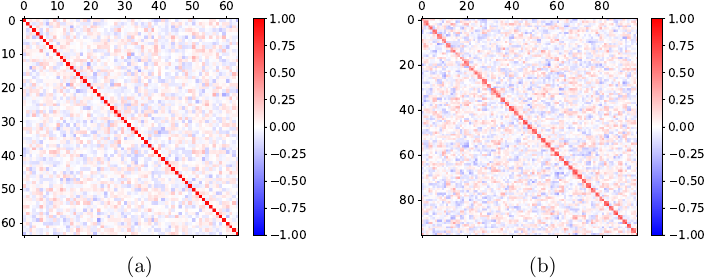}
	\end{centering}
	\caption{\label{fig:gim_corr_mat} The cross-correlation in the Eye diagram between the light-quark and the lightest charm-quark correlation functions for (a) the exploratory study~\cite{PhysRevD.94.114516} at heavier-than-physical light-quark mass and (b) the calculation reported on in this work at physical light-quark mass. Although equal timeslices exhibit a distinguishable correlation in both cases, it is greatly diminished in the physical-point calculation. This results in a poor statistical cancellation in the GIM loop, driving the large statistical error from this calculation.}
\end{figure*}

\section{Conclusion}
We have carried out the first lattice QCD calculation of the $K^{+}\to\pi^{+}\ell^{+}\ell^{-}$ decay amplitude using physical pion and kaon masses. When using physical light-quark masses, even with unphysically light charm-quark masses, the contributions in the GIM loops statistically decorrelate, as shown in Fig.~\ref{fig:gim_corr_mat}. This contributes to the unsatisfactory amount of noise in GIM subtraction, as can be seen in Fig.~\ref{fig:Q1Q2_c1}. Although sparse noises reduced the statistical error introduced by the single-propagator trace contribution to the Eye and Saucer diagrams, we are not able to obtain a well-resolved result for the amplitude.

The form factor that encapsulates the behavior of the long-distance amplitude of the rare kaon decay was found to be $V\left(0.013\left(2\right)\right)=-0.87\left(4.44\right)$. 
When this is compared to experimental results, $V^{exp}\left(0\right) \equiv a^{exp}_{+}=-0.578(16)$ from the electron and $a^{exp}_{+}=-0.592(15)$ from the muon, it can be seen that the error on our lattice result is about $8$ times larger than the central value of the experimental result. However, our error is $3$ times larger than the phenomenological central value obtained in \cite{D_Ambrosio_2019,D_Ambrosio_2019b}, which suggests that lattice QCD calculations will be able to provide a competitive theoretical bound on $a_+$ in the coming years.

We would like to stress that since the noise emerges mainly from the lack of correlation in the GIM subtraction, the error obtained here has the potential to be reduced beyond square-root scaling by optimising the stochastic estimator used for the up-charm loops. Such problems have common elements with similar challenges in computing quark-disconnected diagrams, for example as discussed in~\cite{Giusti:2019kff}. 

Finally, it might also be possible to work in 3-flavor QCD, foregoing the calculation of the charm-quark loop \cite{ALawsonsThesis}, further reducing computational costs. This would require a new renormalization procedure which would be analogous to that of the $K\to\pi\nu\bar{\nu}$ study that was performed by the RBC-UKQCD collaborations previously \cite{Christ_2016_nunubar, Bai_2017_nunubar}. 

In conclusion, despite obtaining a first physical result with a large uncertainty, we believe that optimisation of the methodology, combined with the increased capabilities of future computers, should allow for a competitive prediction of the $K^{+}\to\pi^{+}\ell^{+}\ell^{-}$ amplitude within the next years.

\begin{acknowledgments}
This work used the DiRAC Extreme Scaling service at the University of Edinburgh, operated by the Edinburgh Parallel Computing Centre on behalf of the STFC DiRAC HPC Facility (www.dirac.ac.uk). This equipment was funded by BEIS capital funding via STFC capital grant ST/R00238X/1 and STFC DiRAC Operations grant ST/R001006/1. DiRAC is part of the National e-Infrastructure. PB has been supported in part by the U.S. Department of Energy, Office of Science, Office of Nuclear Physics under the Contract No. DE-SC-0012704 (BNL). FE, VG, R Hodgson, F\'Oh, and AP received funding from the European Research Council (ERC) under the European Union’s Horizon 2020 research and innovation program under grant agreement No 757646 and AP additionally under grant agreement No 813942. R Hill was partially supported by the DISCnet Centre for Doctoral Training (STFC grant ST/P006760/1). AP, VG, FE, and R Hill are additionally supported by UK STFC grant ST/P000630/1. AJ and JF acknowledge funding from STFC consolidated grant ST/P000711/1, and AJ from ST/T000775/1.  CTS was partially supported by an Emeritus Fellowship from the Leverhulme Trust and by STFC (UK) grants ST/P000711/1 and ST/T000775/1.
\end{acknowledgments}

\bibliography{rareKaon2021}

\def\arxiv{1} 

\if\arxiv1
    \newpage
    \setcounter{section}{0}
    \pagenumbering{roman}
    \setcounter{page}{0}
    \setcounter{equation}{0}
    \begin{titlepage}
       \begin{center}
           \vspace*{1cm}
           \textbf{Supplementary material: Simulating rare kaon decays $K^{+}\to\pi^{+}\ell^{+}\ell^{-}$\\using domain wall lattice QCD with physical light quark masses}
       \end{center}
    \end{titlepage}


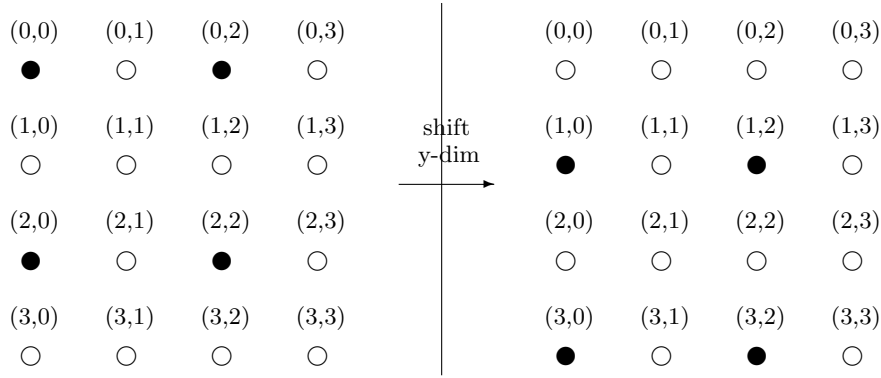
\begin{figure*}
    \begin{minipage}{1.5in}
\setlength{\unitlength}{.05in}
        \begin{picture}(55,40)(0,0)
        \multiput(-33,35)(10,0){1}{\circle*{2}}
        \put(-30,39){\makebox(0,0)[r]{\small{(0,0)}}}
        \multiput(-23,35)(10,0){1}{\circle{2}}
        \put(-20,39){\makebox(0,0)[r]{\small{(0,1)}}}
        \multiput(-13,35)(10,0){1}{\circle*{2}}
        \put(-10,39){\makebox(0,0)[r]{\small{(0,2)}}}
        \multiput(-3,35)(10,0){1}{\circle{2}}
        \put(0,39){\makebox(0,0)[r]{\small{(0,3)}}}
        \multiput(-33,25)(10,0){1}{\circle{2}}
        \put(-30,29){\makebox(0,0)[r]{\small{(1,0)}}}
        \multiput(-23,25)(10,0){1}{\circle{2}}
        \put(-20,29){\makebox(0,0)[r]{\small{(1,1)}}}
        \multiput(-13,25)(10,0){1}{\circle{2}}
        \put(-10,29){\makebox(0,0)[r]{\small{(1,2)}}}
        \multiput(-3,25)(10,0){1}{\circle{2}}
        \put(0,29){\makebox(0,0)[r]{\small{(1,3)}}}
        \multiput(-33,15)(10,0){1}{\circle*{2}}
        \put(-30,19){\makebox(0,0)[r]{\small{(2,0)}}}
        \multiput(-23,15)(10,0){1}{\circle{2}}
        \put(-20,19){\makebox(0,0)[r]{\small{(2,1)}}}
        \multiput(-13,15)(10,0){1}{\circle*{2}}
        \put(-10,19){\makebox(0,0)[r]{\small{(2,2)}}}
        \multiput(-3,15)(10,0){1}{\circle{2}}
        \put(0,19){\makebox(0,0)[r]{\small{(2,3)}}}
        \multiput(-33,5)(10,0){1}{\circle{2}}
        \put(-30,9){\makebox(0,0)[r]{\small{(3,0)}}}
        \multiput(-23,5)(10,0){1}{\circle{2}}
        \put(-20,9){\makebox(0,0)[r]{\small{(3,1)}}}
        \multiput(-13,5)(10,0){1}{\circle{2}}
        \put(-10,9){\makebox(0,0)[r]{\small{(3,2)}}}
        \multiput(-3,5)(10,0){1}{\circle{2}}
        \put(0,9){\makebox(0,0)[r]{\small{(3,3)}}}

        \put(10,3){\line(0,1){39}} 

        \put(5.5,23){\vector(1,0){10}}
        \put(13,29){\makebox(0,0)[r]{shift}}
        \put(14,26){\makebox(0,0)[r]{y-dim}}

        \multiput(23,35)(10,0){1}{\circle{2}}
        \put(26,39){\makebox(0,0)[r]{\small{(0,0)}}}
        \multiput(33,35)(10,0){1}{\circle{2}}
        \put(36,39){\makebox(0,0)[r]{\small{(0,1)}}}
        \multiput(43,35)(10,0){1}{\circle{2}}
        \put(46,39){\makebox(0,0)[r]{\small{(0,2)}}}
        \multiput(53,35)(10,0){1}{\circle{2}}
        \put(56,39){\makebox(0,0)[r]{\small{(0,3)}}}
        \multiput(23,25)(10,0){1}{\circle*{2}}
        \put(26,29){\makebox(0,0)[r]{\small{(1,0)}}}
        \multiput(33,25)(10,0){1}{\circle{2}}
        \put(36,29){\makebox(0,0)[r]{\small{(1,1)}}}
        \multiput(43,25)(10,0){1}{\circle*{2}}
        \put(46,29){\makebox(0,0)[r]{\small{(1,2)}}}
        \multiput(53,25)(10,0){1}{\circle{2}}
        \put(56,29){\makebox(0,0)[r]{\small{(1,3)}}}
        \multiput(23,15)(10,0){1}{\circle{2}}
        \put(26,19){\makebox(0,0)[r]{\small{(2,0)}}}
        \multiput(33,15)(10,0){1}{\circle{2}}
        \put(36,19){\makebox(0,0)[r]{\small{(2,1)}}}
        \multiput(43,15)(10,0){1}{\circle{2}}
        \put(46,19){\makebox(0,0)[r]{\small{(2,2)}}}
        \multiput(53,15)(10,0){1}{\circle{2}}
        \put(56,19){\makebox(0,0)[r]{\small{(2,3)}}}
        \multiput(23,5)(10,0){1}{\circle*{2}}
        \put(26,9){\makebox(0,0)[r]{\small{(3,0)}}}
        \multiput(33,5)(10,0){1}{\circle{2}}
        \put(36,9){\makebox(0,0)[r]{\small{(3,1)}}}
        \multiput(43,5)(10,0){1}{\circle*{2}}
        \put(46,9){\makebox(0,0)[r]{\small{(3,2)}}}
        \multiput(53,5)(10,0){1}{\circle{2}}
        \put(56,9){\makebox(0,0)[r]{\small{(3,3)}}}
        \end{picture}
\end{minipage}
        \caption{An example of sparse noises for $d=2$, $n=2$. The filled circles represent a site with $Z_{2}$ noise, the empty circles represent a site that has been set to zero. Two further shifts are needed to cover the full volume, giving $2^2=4$ sparse sources.}
    \label{fig:sparse_noise_example}
\end{figure*}

\section{\label{sec:sparse_sources}Sparse Sources}
The spacetime distribution of a source may be treated stochastically, in order to decrease the effects of local fluctuations from the gauge fields. This is important for constructing lattice propagators of the form $S\left({x,x}\right)$, which are needed to calculate a disconnected diagram or a single-propagator trace contribution to a correlation function, needed for the Eye and Saucer diagrams (Fig.~\ref{fig:H_W_contractions}) contributing to the rare kaon decay amplitude. To create the propagators we depend on $N$ stochastic sources $\kappa_{i}$ that fulfill the properties
\begin{equation}
\label{eq:stoch_source_props}
\begin{split}
    \lim_{N\to\infty}&\frac{1}{N}\sum_{i=1}^{N} \kappa_{i}\left({x}\right) = 0, \\
    \lim_{N\to\infty}&\frac{1}{N}\sum_{i=1}^{N} \kappa_{i}\left({x}\right)\kappa_{i}^{\dagger}\left({y}\right) = \delta_{xy}.
\end{split}
\end{equation}
One appropriate choice is the $Z_{2}$ source \cite{Z2noise}, where each element is randomly chosen from
\begin{equation}
\label{eq:z2_source_defn}
    Z_{2}\otimes Z_{2} = \left\{\frac{1}{\sqrt{2}}\left({\pm1\pm i}\right)\right\}.
\end{equation}

It is expected that the statistical error introduced from using stochastic sources scales as ${1}/{\sqrt{N}}$. Each stochastic source here covers the full volume but we can also create ``sparse sources'', similar to those described in \cite{PhysRevLett.116.232002}, to improve the ${1}/{\sqrt{N}}$ scaling of the statistical error. In $d$-dimensional spacetime we create $N=n^{d}$ sparse sources where
\begin{equation}
\label{eq:sparse_sources_defn}
    \kappa_{\textrm{sparse}}(x) = 
       \begin{cases}
            \kappa_{Z_{2}}(x):  x_{\mu} ~\mathrm{ mod }~ n = 0, \quad \mu=0,1,2,3 \\
            0: \mathrm{otherwise} \\
        \end{cases}
\end{equation}
for the first source and we shift in each dimension to ensure that the N sources cover the entire volume with no overlap when combined, see Fig.~\ref{fig:sparse_noise_example} for a $d=2$, $n=2$ example. When investigating rare kaon decays we use $N=2^{4}=16$ sparse sources for each hit of a propagator, $S\left({x,x}\right)$, that we compute.\\
A cost-benefit analysis was performed using the quantity $\frac{\Delta X}{\Delta{\textrm{Sparse}}}\sqrt{\frac{N_{X}}{N_{{\textrm{Sparse}}}}}$, where ${\Delta X}$ is the statistical error of the result from method $X$ and the root of the number of inversions $N_{X}$ tracks the computational cost of using method $X$. Fig.~\ref{fig:sparseCost} shows the results of this cost-benefit analysis for the 3-point Saucer diagram with zero momentum. The loop in the diagram was computed using sparse sources, full volume sources and time-diluted all-to-all vectors  \cite{A2APractical} with 2000 low modes, with the other propagators being computed with Coulomb-gauge fixed wall sources. This was performed on RBC/UKQCD's $48^3\times96$ M\"obius domain wall fermion gauge ensembles \cite{PhysRevD.93.074505}. It can clearly be seen that the sparse-noise approach is the most successful.
\begin{figure*}
    \begin{centering}
        \includegraphics[width=0.75\textwidth]{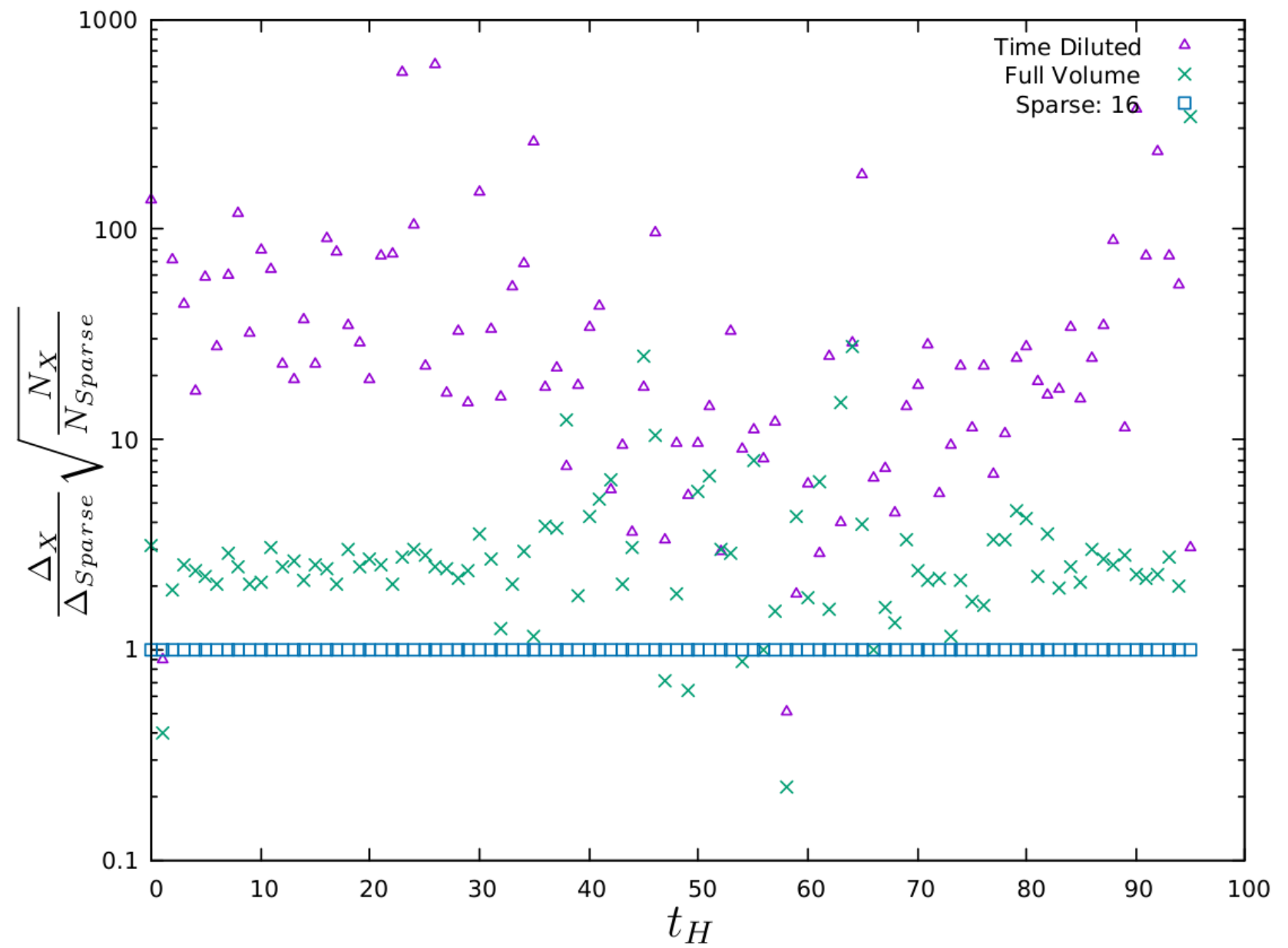}
    \par\end{centering}
\protect\caption{\label{fig:sparseCost}The statistical error relative to that of the ``Sparse: 16'' noise, weighted by the cost of inversions, of the zero-momentum Saucer diagram contribution to the 3-point weak Hamiltonian correlation function, computed using different noise strategies for the single quark propagator loop.}
\end{figure*}

\begin{figure*}
    \begin{centering}
    	\begin{tabular}{cccc}
    		\includegraphics[scale=0.6]{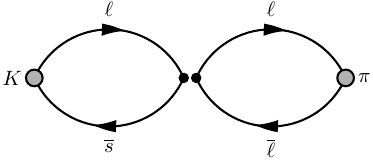} & \includegraphics[scale=0.6]{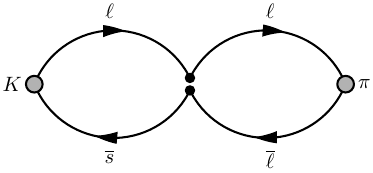} & \includegraphics[scale=0.6]{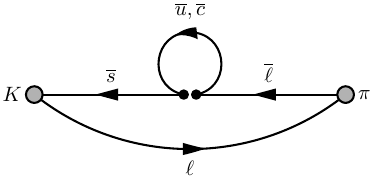} & \includegraphics[scale=0.6]{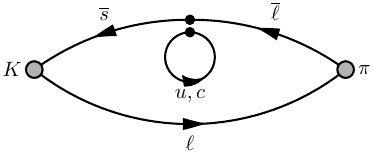}\tabularnewline
    		$W$ & $C$ & $S$ & $E$\tabularnewline
    		(Wing) & (Connected) & (Saucer) & (Eye)\tabularnewline
    	\end{tabular}
	\par\end{centering}
\protect\caption{\label{fig:H_W_contractions}The four classes of diagrams obtained after performing the Wick contractions of the charged pion and kaon interpolating operators with the $H_{W}$ operator. $\ell$ denotes a light ($u$ or $d$) quark propagator. The two black circles represent the currents in the four-quark operators $Q_{1,2}^{q}$ defined in the main paper. 
The C and E diagrams contain an insertion of $Q_{1}^{q}$ and the W and S diagrams contain an insertion of $Q_{2}^{q}$.}
\end{figure*}
\begin{figure}
	\begin{centering}
		\begin{tabular}{cc}
			\includegraphics[scale=0.6]{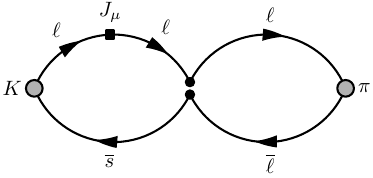} & \includegraphics[scale=0.6]{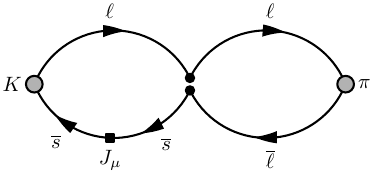} \tabularnewline
			\includegraphics[scale=0.6]{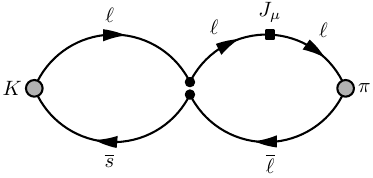} & \includegraphics[scale=0.6]{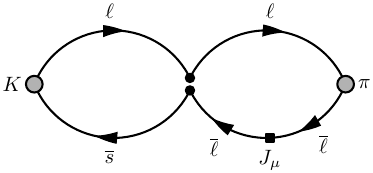} \tabularnewline
		\end{tabular}
			\includegraphics[scale=0.6]{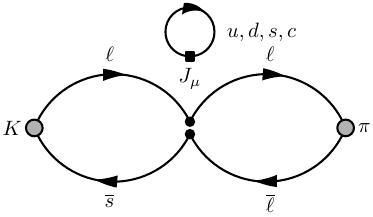}
		\par\end{centering}
	\caption{\label{fig:Connected_J_contractions}The five possible current insertions for the $C$ class of diagrams contributing to the 4pt rare kaon decay correlator. 
	The diagrammatic conventions are the same as in Fig.~\ref{fig:H_W_contractions}}
\end{figure}

\section{\label{relevant_corrs}Further relevant correlators}
Before giving details on the fit parameters that were used we outline the definitions of several relevant Euclidean correlation functions.
\subsection{2-point correlators} 
\label{sub:2point_functions}
Given an interpolation operator $\phi_P(t,\mathbf{p})$ for a pseudoscalar meson, $P$, with spacial momentum $\mathbf{p}$ at time $t$, for $t\gg 0$ the 2-point function
\begin{equation}
    \Gamma^{(2)}_P(t,\mathbf{p})=\langle{\phi_P(t,\mathbf{p})\phi^{\dagger}_P(0,\mathbf{p})}\rangle
\end{equation}
has the following behavior:
\begin{equation}
    \Gamma^{(2)}_P(t,\mathbf{p})=L^{3}\frac{|Z_P(\mathbf{p})|^2}{2E_P(\mathbf{p})}
    [e^{-E_P(\mathbf{p})t} + e^{-E_P(\mathbf{p})(n_t-t)}],
\end{equation}
where $Z_P(\mathbf{p})=\bra{0}\phi_P(0,\mathbf{0})\ket{P(\mathbf{p})}$ and $E_P(\mathbf{p})$ the meson energy $\sqrt{M_P^2+\mathbf{p}^2}$.

We calculated the pion and kaon 2pt functions using Coulomb-gauge fixed wall sources and both Coulomb-gauge fixed wall sinks and point sinks. Although we only require the wall-wall matrix elements in order to extract the decay amplitude the point-wall 2pt functions have a cleaner signal. Thus both the wall-wall and point-wall correlators can be used in a combined fit to obtain $E_P(\mathbf{p})$ with greater accuracy. All pseudoscalar/sink combinations are calculated for $\mathbf{p}=\frac{2\pi}{L}\left(0,0,0\right)$ and $\mathbf{p}=\frac{2\pi}{L}\left(1,0,0\right)$.
\subsection{3-point weak Hamiltonian correlator} 
\label{sub:3_point_h_w}
The weak Hamiltonian 3pt function
\begin{equation}
    \label{eq:3ptHW}
    \Gamma^{(3)}_H(t_H,\mathbf{p})=\int d^3\mathbf{x}\,\langle{\phi_{\pi}(t_{\pi},\mathbf{p})H_W(t_H,\mathbf{x})\phi^{\dagger}_K(0,\mathbf{p})}\rangle
\end{equation}
has the following behavior for $0\ll t_H\ll t_{\pi}$:
\begin{equation}
\begin{split}
    \Gamma^{(3)}_H(t_H,\mathbf{p})=&\,L^3\frac{Z_{\pi}(\mathbf{p})Z_K(\mathbf{p})^{\dagger}\mathcal{M}_H(\mathbf{p})}{4E_{\pi}(\mathbf{p})E_K(\mathbf{p})} \\
                                    & \times e^{-E_{\pi}(\mathbf{p})t_{\pi}}
                                    e^{-[E_K(\mathbf{p})-E_{\pi}(\mathbf{p})]t_H}\,
    \label{eq:GHasymp}
\end{split}
\end{equation} 
with $\mathcal{M}_H(\mathbf{p})=\bra{\pi(\mathbf{p})}H_W(0)\ket{K(\mathbf{p})}$. This correlator is calculated for both $\mathbf{p}=\frac{2\pi}{L}\left(0,0,0\right)$ and $\mathbf{p}=\frac{2\pi}{L}\left(1,0,0\right)$.
\subsection{3-point electromagnetic current correlator} 
\label{sub:3_point_electromagnetic_current}
The electromagnetic current 3pt function for a pseudoscalar meson, $P$,
\begin{equation}
    \Gamma^{(3)\,P}_{J_\mu}(t,t_J,\mathbf{p},\mathbf{k})=\int d^3\mathbf{x}\,
    e^{-i\mathbf{q}\cdotp\mathbf{x}}\,
    \langle{\phi_{P}(t,\mathbf{p})
    J_{\mu}(t_J,\mathbf{x})\phi^{\dagger}_{P}(0,\mathbf{k})}\,\rangle
\end{equation}
has the following asymptotic behavior for
$0\ll t_J\ll t$:
\begin{equation}
\begin{split}
    \Gamma^{(3)\,P}_{J_\mu}(t,t_J,\mathbf{p},\mathbf{k})=&\,L^3
    \frac{Z_P(\mathbf{p})Z_P(\mathbf{k})^{\dagger}\mathcal{M}_{J_\mu}^{P}(\mathbf{p},\mathbf{k})}
    {4E_{P}(\mathbf{p})E_{P}(\mathbf{k})}\\
     &\times e^{-(t-t_J)E_{P}(\mathbf{p})}e^{-t_JE_{P}(\mathbf{k})}\,
    \label{eq:GJasymp}
\end{split}
\end{equation}
where $\mathcal{M}_{J_\mu}^{P}
(\mathbf{p},\mathbf{k})=\bra{P(\mathbf{p})}
J_{\mu}(0)\ket{P(\mathbf{k})}$. This correlator is calculated for both the pion and the kaon.

\section{\label{sec:ama} Variance reduction techniques}

To compute the costly loop diagrams we use a variation of the all-mode-averaging (AMA) technique. On each configuration, for a given operator $\mathcal{O}$, we have an estimator $\mathcal{O}^{\textrm{e}}_{t_j}$ at ``exact" solver precision ($10^{-8}$, $10^{-10}$, $10^{-12}$, and $10^{-14}$ for the light, $c_{1}$, $c_{2}$, and $c_{3}$ quarks, respectively) using $T$ source times $t_j$ and a sparse noise $n_0$. We furthermore have $N$ estimators $\mathcal{O}^{\textrm{ie}}_{t_j, n_i}$ on each source time $t_j$ at ``inexact" solver precision ($10^{-4}$ for all quarks), computed for $N-1$ sparse noise sources $n_i$ in addition to the noise source $n_0$. The AMA estimator we construct from those estimators is
\begin{align}
    \mathcal{O}^{\textrm{zM}}_{t_j} = \left( \mathcal{O}^{\textrm{e}}_{t_j} - \mathcal{O}^{\textrm{ie}}_{t_j, n_0} \right) + \frac{1}{N-1} \sum_{i=1}^{N} \mathcal{O}^{\textrm{ie}}_{t_j, n_i}
    \, ,
\end{align}
where the superscript "zM" highlights that so far, the zM\"obius action has been used.  In a second AMA step, we compute a single estimator $\mathcal{O}^{\textrm{M}}$ for the sparse noise source $n_0$ and source time $t_0$ at ``exact" precision to get our final estimator
\begin{align}
    \mathcal{O} = \left(\mathcal{O}^{\textrm{M}} - \mathcal{O}^{\textrm{zM}}_{t_0}\right) + \frac{1}{T-1} \sum_{j =1}^{T} \mathcal{O}^{\textrm{zM}}_{t_j}
    \, .
\end{align}
The resulting expectation value $\langle \mathcal{O} \rangle = \langle \mathcal{O}^M \rangle$, but at a much reduced variance. In practice, we used $N=10$ and $T=6$, where the 6 source times have been chosen to evenly interlace the time extent of our lattice (16 timeslices apart from each other).

\section{\label{sec:fit_pars}Fit Parameters}%
Results of the decay amplitude are derived from global fits over all correlation functions involved in a specific fit strategy. A summary of the fit parameters for the 2pt and 3pt functions used, consistent across all fit strategies that they enter, is given in Table~\ref{tab:fit_pars}. The range of $T_a$ and $T_b$ used in each case is given in Table~\ref{tab:A0_results-supp}. In Fig.~\ref{fig:corrHeat} we show the correlation matrix of the 2pt and 3pt functions, as well as slices of the integrated 4pt function, which highlights the high degree of correlation between elements of the integrated 4pt function. This correlation structure in the data makes the use of uncorrelated fits for the integrated 4pt function necessary.

\begin{figure*}[t]
    \begin{centering}
        \includegraphics[width=0.75\textwidth]{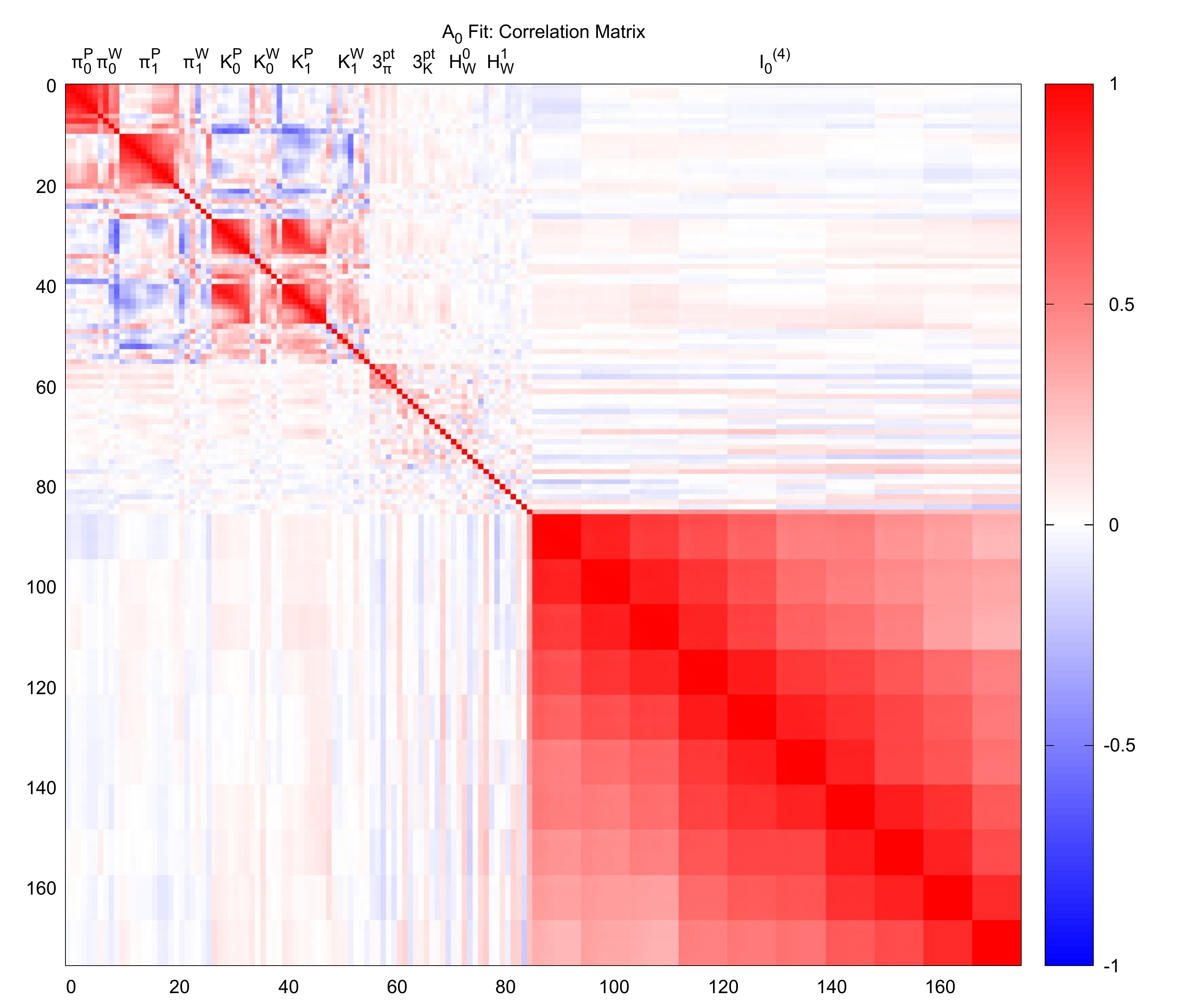}
    \par\end{centering}
\protect\caption{\label{fig:corrHeat}The correlation matrix for the simultaneous fit used to extract $A_{0}$. The correlators in this fit are: the 2pt pion correlator with zero momentum and a point- $(\pi_{0}^{P})$ wall-sink $(\pi_{0}^{W})$, the equivalent correlators with one unit of momentum $(\pi_{1}^{P},~\pi_{1}^{W})$, the same again for the kaon $(K_{0}^{P},~K_{0}^{W},~K_{1}^{P},~K_{1}^{W})$, the vector current inserted on the pion $(3^{pt}_{\pi})$ and kaon $(3^{pt}_{K})$, the 3pt weak Hamiltonian correlator with zero $(H_{W}^{0})$ and one unit of momentum $(H_{W}^{1})$, and the integrated 4pt correlator $(I_{0}^{(4)})$. This shows the correlation matrix for a fully correlated fit. 
Due to the highly correlated nature of the matrix, off-diagonal elements for the integrated 4pt function were set to zero to make the unintegrated 4pt function ``uncorrelated" to the other fit variables.}
\end{figure*}

\begin{table*}[t]
\caption{\label{tab:fit_pars}%
Fit parameters for the various 2pt and 3pt correlators and methods used to extract the $K^{+}\to\pi^{+}\ell^{+}\ell^{-}$ decay amplitude. With the exception of 3pt $c_{s}$ each of these were correlated to each other for the relevant simultaneous fits with the integrated 4pt function. The $c_{s}$ parameter was fitted separately and used as an input to the $c_s$ shift and $c_s\times\bar{s}d$ analyses. The 3pt $H_{W}$ and $c_{s}$ fit parameters are the same for the three unphysical charm-quark masses. ``Thinning" is the stride between data points entering the fit within the fit range.
}
\begin{ruledtabular}
\begin{tabular}{lcccccc}
Correlator&
Momentum&
Sink&
\textrm{$t_{src}$}&
\textrm{$t_{i}$}&
\textrm{$t_{f}$}&
Thinning\\
\colrule
\hline
 2pt pion & $\frac{2\pi}{L}(0,0,0)$ & Point & 32 & 7 & 18  & 2 \\
 2pt pion & $\frac{2\pi}{L}(0,0,0)$ & Wall & 32 & 13 & 20  & 2 \\
 2pt pion & $\frac{2\pi}{L}(1,0,0)$ & Point & 32 & 6 & 25  & 2 \\
 2pt pion & $\frac{2\pi}{L}(1,0,0)$ & Wall & 32 & 9 & 22  & 2 \\
 2pt kaon & $\frac{2\pi}{L}(0,0,0)$ & Point & 0 & 10 & 23  & 2 \\
 2pt kaon & $\frac{2\pi}{L}(0,0,0)$ & Wall & 0 & 9 & 20  & 2 \\
 2pt kaon & $\frac{2\pi}{L}(1,0,0)$ & Point & 0 & 11 & 26  & 2 \\
 2pt kaon & $\frac{2\pi}{L}(1,0,0)$ & Wall & 0 & 10 & 25  & 2 \\
 3pt pion & - & Wall & 0 & 1 & 10  & 2 \\
 3pt kaon & - & Wall & 0 & 20 & 35  & 2 \\
 3pt $c_{s}$ & $\frac{2\pi}{L}(0,0,0)$ & Wall & 0 & 9 & 24  & 1 \\
 3pt $H_{W}$ & $\frac{2\pi}{L}(0,0,0)$ & Wall & 0 & 17 & 24  & 2 \\
 3pt $H_{W}$ & $\frac{2\pi}{L}(1,0,0)$ & Wall & 0 & 13 & 17  & 1 \\
\hline
\end{tabular}
\end{ruledtabular}
\end{table*}

\begin{table*}[t]
\caption{\label{tab:A0_results-supp}%
Fit parameters for the various methods of extracting $A_{0}$. No thinning was performed on the integrated 4pt function data---meaning that all data points within the fit range entered the fit---which were uncorrelated when fitted simultaneously with the relevant 2pt and 3pt correlators.
}
\begin{ruledtabular}
\begin{tabular}{lcccc}
Analysis&
\textrm{$T_{a}$ min}&
\textrm{$T_{a}$ max}&
\textrm{$T_{b}$ min}&
\textrm{$T_{b}$ max}\\
\colrule
\hline
 Direct fit & 2  & 10  & 4 & 11 \\
\hline
 2pt/3pt recon & 1 & 10 & 7 & 15 \\
\hline
 0 mom transfer & 1 & 12 & 3 & 8 \\
\hline
 $SU(3)$ symm lim & 1  & 10  & 6 & 13 \\
\hline
 $c_s$ shift & 1 & 8 & 5 & 12 \\
\end{tabular}
\end{ruledtabular}
\end{table*}

\clearpage


\fi
\end{document}